\documentclass[twocolumn,showpacs,preprintnumbers,amsmath,amssymb,superscriptaddress,prb]{revtex4}
\usepackage{graphicx}
\usepackage{graphics}
\usepackage[dvips,usenames]{color}
\usepackage{dcolumn}
\usepackage{bm}
\usepackage{draftcopy}

\newcommand{\nl}{\newline}
\newcommand{\nn}{$\langle$$n$$\rangle$}

\begin{document}
\title{The influence of proximity effects in inhomogeneous electronic states}


\author{Matthias Mayr}
\affiliation{Department of Physics and Astronomy, The University of Tennessee, Knoxville, Tennessee 37996 and 
Condensed Matter Sciences Division, Oak Ridge National Laboratory, Oak Ridge, Tennessee 37831}

\begin{abstract}

We describe the high-temperature superconductor La$_{2-x}$Sr$_x$CuO$_4$ in the underdoped regime in terms of a mixed-phase 
model, which consists of superconducting clusters embedded in the antiferromagnetic matrix. 
Our work is motivated by a series of recent angle-resolved photoemission experiments, which have significantly enhanced our 
understanding of the electronic structure in single-layer cuprates, and we show that those 
results can be {\it fully} reproduced once a reasonable set of parameters is chosen and disorder is properly taken into account. 
Close to half-filling, many nominally hole-rich and superconducting sites have comparatively large excitation gaps due to 
the ubiquitous proximity of the insulating phase. No other competing states or any further assumptions are necessary to account for a satisfying 
description of the underdoped phase, including the notorious pseudogap, which emerges as a natural consequence of a mixed-phase state. 
Close to optimal doping, the resulting gap distribution mirrors the one found in tunneling experiments.
The scenario presented is also compelling because the existence of an optimal doping is intrinsically linked 
to the presence of electronic inhomogeneity, as is the transition from the highly mysterious to the rather conventional nature of the 
metallic state above $T_c$ in the overdoped regime. The phase diagram that emerges bears strong similarity to the canonical cuprate phase diagram, 
it explains quantitatively the rapid loss of long-range antiferromagnetic order, and also provides for a ``superconducting fluctuating'' regime 
just above $T_c$ as recently proposed in the context of resistance measurements. Implications for manganites are briefly discussed.    

\end{abstract}
\maketitle

\section{Introduction}
After almost two decades of intense research, many fundamental issues in high-temperature superconductivity remain 
unresolved \cite{Dagotto_0}. Besides from elucidating the mechanism responsible for pairing - although antiferromagnetic (AF) spin fluctuations 
have emerged as a strong favorite \cite{Borisenko_1} - 
there is still no satisfying understanding of the particularly ``tricky'' underdoped phase, which is most of all characterized by the notorious pseudogap (PG) 
in the electronic excitation spectrum. Recently, remarkable experiments have presented a much clearer view of this regime, in particular 
angle-resolved photoemission (ARPES) in La$_{2-x}$Sr$_x$CuO$_4$ (LSCO)\cite{Yoshida_1} as well as in Ca$_{2-x}$Na$_x$CuO$_2$Cl$_2$ (Na-CCOC) 
\cite{Kohsaka_1} and scanning tunneling microscopy (STM) in Bi$_2$Sr$_2$CaCu$_2$O$_{8+\delta}$ (Bi-2212)\cite{Lang_1, McElroy_0, Vershinin_1, Howald_1, Mashima_1}, 
the latter revealing a granular, inhomogeneous electronic structure on a nanometer scale. 
Those two experimental techniques would neatly complement each other - 
one mapping out the electronic structure in momentum space, the other one in real space - were it not for the inconvenience that they 
cannot be performed on the same compounds and/or doping levels. Here we want to focus on the interpretation of ARPES experiments, 
since the corresponding numerical calculations are not as much affected by finite-size effects than those with regard to STM data. 
The ARPES data themselves were shown to have provided strong evidence towards the description of the underdoped state in terms of a spatially 
inhomogeneous mixture of hole-rich (i.e., SC) and hole-poor (i.e., AF) clusters\cite{Alvarez_1}. These conclusions are backed up by a variety of other, 
unrelated, experiments such as Raman scattering \cite{Machtoub_1} and $\mu$SR-spectroscopy in the same reference compounds (LSCO), 
as well as by transport measurements in LSCO and YBa$_2$Cu$_3$O$_y$\cite{Padilla_1}, amongst others.

Most importantly, the aforementioned photoemission (PE) data have granted a unique insight into the electronic structure 
of LSCO over the {\it whole} doping range \cite{Damascelli_1}, 
and have elegantly highlighted the fundamental differences between the underdoped and the overdoped regime. Whereas the latter 
shows fairly conventional behavior - comparably to a common, phonon-mediated, superconductor - the underdoped phase shows dramatic deviations 
from the usual Fermi-liquid physics even in the SC state: the spectral intensity was found to have a two-peak structure in the anti-nodal direction, 
across the whole range from zero to optimal doping, $x_{\rm c}$. One of those peaks occupies the same energy position (-0.6eV) as in the undoped Mott insulator 
and thus can safely be identified with the lower Hubbard band (LHB), whereas the second one appears much closer to the Fermi level $E_{\rm F}$ and does so only for 
finite doping concentrations. In addition, finite (although relatively weak) spectral weight along the BZ diagonal exists even deep in the insulating phase ($x$=0.03), 
suggesting that the Mott insulating phase contains aspects of metallicity. Very recently, vastly improved PE data have become available \cite{Yoshida_2}. 
Although they confirm that the second anti-nodal peak in the spectral density distribution is found at (binding energy) $\omega$$\sim$-0.2eV, the accompanying 
distribution is a very flat one, and some weight appears close to $E_{\rm F}$, precisely what should be expected 
for a sufficiently disordered system. Those measurements also confirmed that the main intensity peak itself travels towards $E_{\rm F}$ as more charge carriers are added, 
and eventually assumes more quasi-particle (qp) characteristics, i.e., the associated spectral weight distribution becomes sharper. 

Overall, the transition from the underdoped to the overdoped regime appears to be a smooth one, and seems facilitated by the ``fading away'' 
of the LHB peak and the concurrent strengthening of the spectral intensity close to $E_{F}$, both in the nodal and in the anti-nodal direction. 
As was demonstrated before \cite{Mayr_1}, most of those results can be understood in a rather simple framework of competing AF 
and SC clusters. Doping therein amounts to shifting the relative weight of either phase; at optimal doping AF clusters will have vanished, 
leaving behind a homogeneous SC phase. It was also remarked that the inhomogeneous state in underdoped LSCO has features of a ``ghost'' FS in the sense that 
there exists zero-energy spectral weight for anti-nodal momenta as well. Those data, once combined with the ones from the BZ diagonal, allow for a reconstruction of the 
underlying FS in both theory and experiment. In fact, it was shown for LSCO - but not for Na-CCOC - that a FS constructed in this way even fulfills 
Luttinger's theorem \cite{Yoshida_2}.
One of the main goals here is to expand on our previous calculations by investigating the temperature dependence of the mixed-state as well as to explore 
the corresponding phase diagram. So far, LSCO ARPES data have only been provided for a fixed temperature of $T$=20K. Higher temperatures are difficult to achieve 
for technical reasons, but nonetheless, particularly in the underdoped regime, one can make predictions about the mixed AF/SC state that 
should be experimentally verifiable.\nl   

A quite puzzling result, however, is the development of the doping-induced signal with its observed shift from 0.2eV below $E_{\rm F}$ to -0.03eV 
in the anti-nodal direction. The latter value is certainly in line with what one would expect for a superconductor with a $T_c$ of about 35K as is LSCO, whereas 
the former one stands in stark contrast to the small transition temperatures (actually $T_c$=0) in this regime of the LSCO phase diagram.   
This curious aspect has not been reproduced or even investigated in depth in the previous effort, 
although some possible explanations were suggested to explain the apparent large excitation gap in the very underdoped regime. 
The simplest one is that the superconductor enters a strongly coupled regime as LSCO becomes more underdoped, with the effective qp attraction growing. 
Nevertheless, this is more of a restatement of the 
observed phenomenon, and not an explanation per se. Although it seems conceivable that the charge carriers lose kinetic energy as 
the superconductor approaches the insulating limit, this viewpoint is difficult to reconcile with the presented scenario, because 
in the strong-coupling picture doping leads to a global loss in correlation energy, and not to a local change by replacing insulating sites 
with metallic ones as proposed here.
The strong-coupling scenario has been amply discussed in attempts to explain the PG in terms of preformed pairs and (classical) phase fluctuations, which are of 
fundamental importance for a strongly coupled two-dimensional (2D) superconductor. However, it needs to be noted that such thermal phase fluctuations can in fact be studied 
by Monte-Carlo (MC) techniques without any further approximations\cite{Mayr_2}, and the results suggest that the increase in the excitation gap is not 
a strongly coupled phenomenon: Since even the very underdoped cuprates have spectral weight at $E_{\rm F}$ along the BZ diagonal, 
whereas a superconductor with nearest-neighbor (n.n.) 
attraction undergoes a change in symmetry from $d$-wave to $s$+i$d$ for a sufficiently strong interaction, and which in turn is observable by a full gap in the 
excitation spectrum, LSCO apparently does not enter a strongly-coupled regime.

Alternatively, one may assume that at least one other order parameter (OP) is present that has not yet been accounted 
for in the previous considerations and which would contribute to and enhance the SC gap. One of the best known ideas in this this context is the 
$d$-density wave (DDW) \cite{Chakravarty_1}. Since it is well known that a n.n. attraction may give rise to instabilities other than $d$-wave superconductivity, it is not 
unreasonable to assume that other orderings may be present as well. The best candidates are charge density waves (CDW), which are well known to compete 
with superconductivity in 
n.n. models. The DDW is a Cooper-pair CDW, and, as proposed by  Chakravarty {\it et al.}, competes with the $d$-SC phase in the same way as the insulating phase. 
Of course, it 
also has the same momentum dependence as the $d$-SC, and could appear just as such, if experiments not susceptible to magnetic effects are performed.   
Following the spectroscopic data, however, the DDW OP must exist {\it 'on top'} of the superconductor (i.e., the hole-rich clusters have both a finite DDW as 
well as $d$-SC OP) and 
weaken with the addition of holes although the local density stays the same. There is certainly no compelling reason why this ought to be the case. 
Here, we will propose a much simpler explanation, with the particular advantage that no additional degrees of freedom or exotic phases need to be 
introduced and the desired effect will simply result as a consequence of mixing two different electronic orderings with vastly different energy scales.   

This paper is structured as follows: In Section II, we introduce the Hamiltonians and the mean-field method that we use to obtain the results in later sections.
Then, in Section III we discuss the significance of chemical potential pinning and contrast that with some calculations where disorder effects have been discarded, 
and fluctuating stripes emerge as the hallmarks of doped Mott-Hubbard (MH) insulators. In Section IV the spectral function for mixed AF/SC systems is calculated, 
and in particular we focus on its temperature dependence, to complement the previous effort, where only the ground state has been studied. 
In this section, we will also take a closer look at the excitation gap in the underdoped regime and suggest a simple explanation for the 
observed shift of the low-energy band in the anti-nodal direction by showing that this is a consequence of the spatially inhomogeneous AF/SC mixture. 
In the following Section (V), we discuss the percolative aspects of the SC/non-SC transition and present the phase 
diagram as calculated within the mixed-state scenario and stress its similarity with the canonical cuprate phase diagram. 
The conclusions, Section VI, summarize our findings and provide an outlook for future work, especially with regard to manganites.       
\section{The Hamiltonian and the Self-consistent Solution}
Details of the calculations have already been provided in the previous work, so here we just repeat the essential steps and point out some differences as we consider 
finite temperatures. We investigate the effective mean-field Hamiltonian $H_{\bf HF}$,
\begin{eqnarray}\nonumber
H_{\bf HF}&=&-t\sum_{<{\bf ij}>,\sigma}(c^\dagger_{{\bf i}\sigma}c^{}_{{\bf j}\sigma}
+H.c.) -\sum_{{\bf i}\sigma} \mu_{\bf i} n_{{\bf i}\sigma} \\ 
& - & 
\sum_{{\bf \langle ij \rangle}}(\Delta_{{\bf ij}}
c_{{\bf i}\uparrow}c_{{\bf j}\downarrow}+H.c.) + \frac{1}{2}\sum_{\bf i} U_{\bf i} m_{\bf i} s_{\bf i}^z + \nonumber \\
& + & \sum_{{\bf \langle ij \rangle}}V_{\bf ij}|\Delta_{\bf ij}|^2 +\- 1/4 \sum_{\bf i}U_{\bf i} (\langle n_{\bf i}\rangle^2-m^2_{\bf i}), 
\label{eq:hamfermi0}
\end{eqnarray} 
that can be derived from the original extended Hubbard model in a Hartree-Fock approximation, assuming both finite SC and magnetic order exist in certain parts of the sample. 
$c^{\dagger}_{{\bf i}\sigma}$ in Eq.(\ref{eq:hamfermi0}) are electron (creation) operators on a two-dimensional quadratic lattice of $N$=$L$$\times$$L$ sites, 
and $t$ is their concomitant hopping amplitude between n.n. sites ${\bf i}$, ${\bf j}$. We chose $t$=1 as the energy unit. The particle density 
$n_{\bf i}$=$\sum_{\sigma}$$c^\dagger_{{\bf i}\sigma}$$c^{}_{{\bf i}\sigma}$ is determined by the chemical potential $\mu_{\bf i}$ and both are assumed to vary spatially. 
In Eq.(\ref{eq:hamfermi0}), we have also introduced the local spin operator $s^z_{\bf i}$=$\frac{1}{2}$($n_{{\bf i}\uparrow}$-$n_{{\bf i}\downarrow}$). 
The local magnetic OP $m_{\bf i}$ and the site-dependent ($d$-wave) SC OP $\Delta_{\bf ij}$ are related to the fermion operators,
\begin{eqnarray}\nonumber
m_{\bf i} & \equiv & \langle n_{\bf i\uparrow} \rangle - \langle n_{\bf i\downarrow} \rangle\\
\Delta_{\bf ij} & \equiv & \hspace*{0.3cm} V_{\bf ij} \langle c_{\bf i\uparrow}c_{\bf j\downarrow}\rangle,
\label{eq:orderpara}
\end{eqnarray} 
where $\langle$$\hdots$$\rangle$ denotes expectation values after thermal averaging. $V_{\bf ij}$ (defined on links ${\bf i,j}$) is the n.n. attraction 
that appears in the original extended Hubbard Hamiltonian and 
it is responsible for the occurrence of SC (or charge-order (CO), if a charge-density wave were to be considered), whereas $U_{\bf i}$ is the on-site Coulomb interaction, 
leading to AF order. The Hamiltonian $H_{\bf HF}$=$H_{\bf HF}'$+$H^{cl}_{\bf HF}$ is quadratic in electron operators, which makes it 
a single-particle Hamiltonian with 2$N$ basis states, that can be readily diagonalized 
using library subroutines. $H^{cl}_{\bf HF}$ (third line in Eq.(\ref{eq:hamfermi0})) is a solely classical energy term, with no operators involved. It simply changes the 
configuration energy, but will not be of importance for the rest of the paper. Eqs. (\ref{eq:hamfermi0}), (\ref{eq:orderpara}) constitute a self-consistent system, 
which can be solved iteratively. This amounts to performing a Bogoliubov-de Gennes (BdG) transformation 
\cite{Atkinson03, Ghosal_1, Ichioka_1}, whereby the original operators $c_{{\bf i}\sigma}$ are expressed in terms of new Bogoliubov 
quasiparticles $\gamma_{n\sigma}$:
\begin{eqnarray}
c_{\mathbf i\uparrow}&=&\sum_{n=1}^{N}\{ a_n(\mathbf i)\gamma_{n\uparrow}-
b^{*}_{n+N}(\mathbf i)\gamma_{n\downarrow}^\dagger\},\nonumber\\ 
c_{\mathbf i\downarrow}&=&\sum_{n=1}^{N}\{ b_n(\mathbf i) \gamma_{n\downarrow}+
a^{*}_{n+N}(\mathbf i)\gamma_{n\uparrow}^\dagger\}.
\label{eq:bogoliubov}
\end{eqnarray}
This, together with Eq. (\ref{eq:orderpara}), leads to the following expressions for the OP's as functions of the wave functions $a_n$(${\bf i}$), $b_n$(${\bf i}$):
\begin{eqnarray}\nonumber
\Delta_{\bf ij} &=&\sum^{N}_{n=1} a^{}_{n}({\bf i})a^*_{n+N}({\bf j})\langle \gamma^{}_{n\uparrow}\gamma^{\dagger}_{n\uparrow}\rangle - b^*_{n+N}({\bf i})b_n({\bf j})\langle \gamma^{\dagger}_{n\downarrow}\gamma^{}_{n\downarrow}\rangle,\\
n_{{\bf i}\uparrow}&=& \sum^N_{n=1}|a_n({\bf i})|^2\langle \gamma^{\dagger}_{n\uparrow}\gamma^{}_{n\uparrow}\rangle + |b^{}_{n+N}({\bf i})|^2\langle \gamma^{}_{n\downarrow}\gamma^{\dagger}_{n\downarrow}\rangle, \nonumber \\
n_{{\bf i}\downarrow}&=& \sum^N_{n=1}|b_n({\bf i})|^2\langle \gamma^{\dagger}_{n\downarrow}\gamma^{}_{n\uparrow}\rangle + |a^{}_{n+N}({\bf i})|^2\langle \gamma^{}_{n\downarrow}\gamma^{\dagger}_{n\downarrow}\rangle.
\label{eq:self}
\end{eqnarray} 
where $\langle \gamma^{\dagger}_{n\sigma}\gamma^{}_{n\sigma} \rangle$=$\{ 1+e^{\beta E_{n\sigma}}\}^{-1}$, 
i.e. the Fermi function $f$ with properties $f$($x$)=1-$f$(-$x$), $\beta$=1/$T$ the inverse temperature and $E_{n\sigma}$ 
are the eigenvalues of Eq.(\ref{eq:hamfermi0}). Whereas previously we had focused on ground state properties only, this time we will investigate specifically 
the temperature dependence. At the start of the iteration initial values $\Delta^0_{\bf ij}$, $m^0_{\bf i}$ are chosen, and the resulting 
Hamiltonian Eq. (\ref{eq:hamfermi0}) is diagonalized, leading to wave functions $a^0_n$(${\bf i}$), $b^0_n$(${\bf i}$) and eigenvalues $E^0_{n\sigma}$. 
Those are then used to compute $\Delta^1_{\bf ij}$, $m^1_{\bf i}$ 
via Eq.(\ref{eq:self}), and so forth. Those iterations are stopped once $\frac{1}{N}$$\sum_{\bf ij}$$|$$\Delta^n_{\bf ij}$-$\Delta^{n-1}_{\bf ij}$$|$$<$$\epsilon$, 
where $\epsilon$ typically is $10^{-5}$, and a similar condition applies to the magnetic OP as well. In case no magnetic order is present $a_n$$\equiv$$b_n$, 
leading us back to the original BdG equations. 

Using mean-field approximations, it is very well known that for Hamiltonians such as $H_{\bf HF}$ charge-inhomogeneous regions 
appear {\it spontaneously} in the low-doping limit 
($\delta$=1-$\frac{1}{N}$$\sum_{\bf i}n_{\bf i}$, $\delta$$\ll$1) in the form of stripes \cite{Zaanen_1}. This regime is difficult to handle numerically, 
and may not even fully reflect the physics of underdoped cuprates since it neglects the 
influence of chemical disorder. Naturally, one would expect such terms due to the presence of the Mott insulating phase and its associated poor screening 
properties. Therefore, in our case doping is always accompanied by a change in the local chemical potential $\mu_{\bf i}$, which is set so that the 
resulting local charge density $n_{\bf i}$ is the one of optimal doping. This is achieved by introducing ``plaquettes'', which are small patches of reduced chemical potential 
(and thus charge) and enhanced pairing \cite{Wang_1}. This expresses the view that Coulomb potentials, maybe in conjunction with the AF background,
 partially bind holes to their donor. Then, the chemical potential is composed of a local term {\it and} a global 
one, $\mu_0$; the latter still has to be determined 
self-consistently to obtain the desired global charge (or hole) density. In order for our results to reflect the PE data, we will compare the computed values 
of $\mu_0$ with the ones obtained via PE in LSCO.        

From the knowledge of the eigenfunctions, all relevant observables such as the single-particle spectral function 
\begin{equation}
A({\bf k}, \omega) = -\frac{1}{\pi}Im G({\mathbf k}, \omega),
\end{equation} 
$G({\mathbf k}, \omega)$=$\sum_{{\bf i,i'},\sigma}e^{i{\bf k}({\bf i}-{\bf i'})}\langle c^{}_{{\bf i}\sigma}c^\dagger_{{\bf i'}\sigma}$$\rangle$ 
being the usual Green's function, and the local density of states 
\begin{equation} 
N({\bf i}, \omega)  = -\frac{1}{\pi}Im G({\mathbf i}, \omega)
\end{equation}
can be 
calculated in a straightforward fashion. The density of states $N$($\omega$)=$\frac{1}{N}$$\sum_{\bf i}$$N({\bf i}, \omega)$ follows immediately. 
Those observables are the crucial quantities with regards to ARPES and STM experiments, respectively. 
In the two limiting clean cases, $\langle$$n$$\rangle$=1 and with no SC regions present, as well as $\langle$$n$$\rangle$=0.75 ($\equiv$$n_{\rm SC}$) 
and without residual AF islands, the BdG equations converge fairly quickly and one obtains the curves $\Delta$($T$) and $m$($T$) 
(spatial indices are not necessary since the systems are homogeneous), 
which can be compared with 
results from solving the self-consistent equations in momentum-space. The temperatures $T_c$ upon where $\Delta$($T_c$)=0 ($m$($T_c$)=0) 
define the corresponding critical temperatures. For the two cases considered, one finds $T^{AF}_c$$\sim$1 and $T^{SC}_c$$\sim$0.35 ($\langle$$n$$\rangle$=0.75) 
for the parameter values $U$=5 and $V_{\bf ij}$=$V$=1, 
respectively, which we will adopt from here on, unless mentioned otherwise. It is mostly important for our purposes that they are sufficiently separated, but we do point out 
here that the separation of energy scales is not as large as in LSCO, where $T^{AF}_c$$\approx$10$T^{SC}_c$, an issue we will return to later.   

\section{The chemical potential and the influence of disorder}
As mentioned above, the local chemical potential is chosen so that the resulting charge density at hole-rich sites stays approximately constant, and further 
doping is mimicked by introducing more such charge-depleted regions. Effectively, as we will also show below, this results in a pinning of 
the global chemical potential $\mu_0$ at the original Fermi level. Since this pinning, according to ARPES, holds up 
to roughly optimal doping upon where a steep drop in $\mu_0$ in agreement with common band-structure theory is registered, it is natural to assume that the 
chosen charge density level for hole-rich clusters should be the one of optimal doping and we adjust $\mu_{\bf i}$ accordingly. 

\begin{figure}
\includegraphics[width=7.0cm,clip]{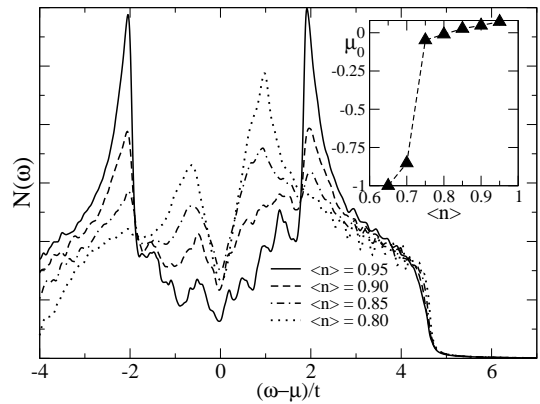}
\caption{\label{fig:dos} Density of states $N$($\omega$) for the mixed SC/AF system on a 40$\times$40 lattice at low temperatures. 
Parameters are given in the text. The low-energy peak at small $x$ is not intrinsically connected to the prominent coherence peak 
observed for $x$=0.20, but is an accidental consequence of the plaquette structure chosen. The inset shows the chemical potential, 
$\mu_0$, as a function of the overall density. $\mu_0$ is almost completely pinned to 
the original Fermi level $\omega$=0 until the mixed-phase regime is replaced by a homogeneous (SC) state, 
upon where $\mu_0$ drops in accordance with band-theory.}
\end{figure}

Fig.\ref{fig:dos} shows the resulting low-temperature $N$($\omega$) as a function of density, obtained after averaging over 3 configurations 
on a 40$\times$40 lattice. Given the comparatively large size of our lattice, even such a relatively small number should be sufficient to 
guarantee configuration-independent results. By comparing results obtained for individual configurations we also have confirmed that 
there are only minor variations in between them and the systems therefore are close to self-averaging. The large spike associated with 
the MH insulator appears at $\omega_{\rm MH}$$\sim$-2, and is determined by our choice of $U$=5. 
Upon doping, in-gap (i.e., between this MH gap) states are created, and they are continuously filled as the system evolves towards $x_{\rm c}$. 
The MH peak in $N$($\omega$) diminishes in size, and is eventually replaced by the SC coherence peak positioned at $\omega$$\simeq$-0.7. 
Nevertheless, as the original gap is filled by the hole-rich regions, the spike at $\omega_{\rm MH}$ remains in its original position throughout 
the doping process. This is the LHB pinning observed in ARPES. The small signal at $\omega_{\rm MH}$/3 ($\langle$$n$$\rangle$=0.95) 
corresponds to the second band found along the anti-nodal direction upon doping (see Refs. \onlinecite{Yoshida_1, Mayr_1}). Whereas in PE, however, 
this peak moves towards $E_{\rm F}$, in Fig.\ref{fig:dos} the opposite behavior is observed. This issue has caused some confusion in the previous work \cite{Mayr_1}, 
and it will be discussed in great detail in a later section.

The inset in Fig.\ref{fig:dos} shows how $\mu_0$ develops: it remains pinned at the original level $\omega$=0 for densities down to $n_{\rm SC}$, 
but drops steeply once the system has become homogeneous. 
Although this is congruent with what has been found in LSCO, by no means does this describe all underdoped cuprates as a more conventional 
behavior is found for Na-CCOC, where the (surprisingly) broad band associated with the AF insulator moves continuously towards $E_{\bf F}$ with the 
addition of holes. It is also important to consider that the scenario presented here could still be somewhat oversimplified, since a careful analysis of spectral data 
seems to suggest that although an intermediate band is indeed created upon doping, the chemical potential does move slightly downward {\it inside this band} as more 
holes are added. Whether the observed small shift in $\mu_0$ between $x$=0 and $x$=0.25 (see inset Fig.\ref{fig:dos}) describes the same effect is, 
as of now, not clear, but we will further comment on this below. 

The importance of disorder and its correlations with the electronic structure were recently demonstrated in STM investigations of slightly underdoped 
Bi-2212 \cite{McElroy_1}, and the influence of dopant atoms was related to a change of the local chemical potential as far as the CuO-layers are concerned. 
There, it was also remarked that the charge fluctuations are surprisingly small, of the order of 10$\%$. This is basically the viewpoint taken here as well. 
Alternatively, the case where doping is not accompanied by disorder should also be considered, 
since there are indications that not all high-$T_c$ compounds are equally disordered. Although in principle doping without accompanying change in the 
local chemical potential $\mu_{\bf i}$ (as described in Section III) can be achieved by simply reducing $\mu_0$, in practice this is not possible 
within the mean-field framework employed here. 
As is well known, Eq.(\ref{eq:hamfermi0}) is unstable towards phase-separation in the form of stripes, and the intermediate regime between 
the Mott insulator at $\langle$$n$$\rangle$=1 and the stripe phase at a much smaller, precisely determined, value of $\langle$$n$$\rangle$ cannot be accessed 
(in particular for the lattice sizes considered). Yet, as it turns out, this phase can be adequately described 
via a MC technique that was introduced in the previous paper\cite{Mayr_1}, albeit at the price of being limited to smaller lattices.  
\begin{figure}
\includegraphics[width=7.0cm,clip]{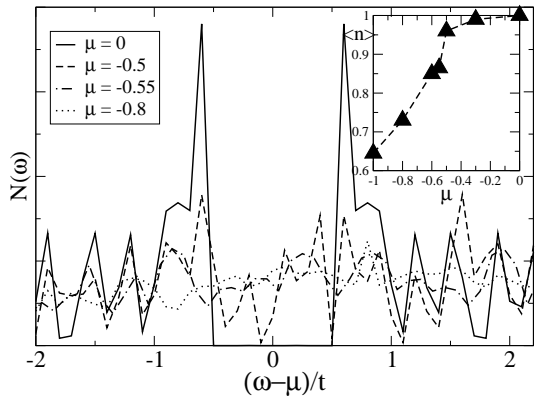}
\caption{$N$($\omega$) derived from a MC integration of $H_{\bf HF}$ on a 16$\times$16 lattice. $J$$\equiv$$\frac{1}{2}$$Um$=0.6 and $T$=0.04. 
Note that these are original data without Lorentzian broadening. The inset shows the particle density as a function of the chemical potential.}
\label{fig:mc_nw}
\end{figure}
$N$($\omega$) resulting from a MC integration of $H_{\bf HF}$ at low, but finite, temperature is displayed in Fig.\ref{fig:mc_nw} 
for a relatively large lattice of $N$=16$\times$16 sites. For this calculation spatial variations of $\mu_{\bf i}$ and $U_{\bf i}$ were not considered. 
Doping with charge carriers is mimicked by reducing the (global) chemical potential, as underlined in the inset of Fig.\ref{fig:mc_nw}:  
As $\mu$ is reduced, the density drops concurrently. This process, however, may not be entirely smooth: in fact, a tendency to phase separation is found 
around $\mu$=-0.5, between densities $\langle$$n$$\rangle$=0.94 and $\langle$$n$$\rangle$$\simeq$0.87 ($x$$\sim$1/8), signaling a first order phase transition. 
Similar results were also found for models that implement the SO(3) symmetries of the classical spins \cite{Buehler_1}, which shows that the 
Ising spin approximation employed here is not a crucial one. Even in the slightly doped case, the MC-averaged density is constant throughout the lattice, 
but snapshots show that single sites are either almost perfectly half-filled or, to a much smaller degree, possess a density \nn$\sim$0.63, 
this value mainly being a function of the magnetic coupling $J$=$\frac{1}{2}Um$, which, instead of $U$, enters the MC process \cite{Comment_1}. 
Hole-rich sites sometimes arrange themselves along straight or diagonal lines, without forming full stripes. Essentially, the MC process describes a phase 
of {\it dynamical} stripes, which are broken in pieces either because of thermal fluctuations or because the densities involved do not allow for a full stripe 
to be formed. Notably, in this regime the original spike in $N$($\omega$) at energy $J$ remains at its position (Fig.\ref{fig:mc_nw}, long-dashed line) and is 
only diminished in size, whereas at the same time finite spectral weight appears at low-energies, $\omega$$\sim$$J$/3-$J$/2. One peak clearly relates to the
fully occupied sites, whereas the latter one reflects the residual magnetic gap at hole-rich sites, a view underscored by a direct measurement 
and also by the observation that its position scales with $J$ over a range of $J$ (0.3$\leq$$J$$\leq$0.8) values observed. This is quite similar to what is 
found in Fig.\ref{fig:dos}, but upon further doping this two-peak structure is lost, and replaced by a smooth function without any noticable features 
in $N$($\omega$), even though charge-disorder prevails. For example, even 
at the stripe-favoring hole concentration level $x$$\simeq$1/8, do we not find the naively expected stripe state, 
but instead stripe-like states with charge distributions centered around densities $n_1$$\sim$0.94, $n_2$$\sim$0.75, 
and similar results for other values of $\mu$. 
In other words, we have disorder without pinning. This stands in contrast to LSCO PE data, where pinning and finite excitation gaps are observed across the 
underdoped regime. Similar results were found for a wide variety of values of $J$, and thus they should be regarded as representative for disorder-free 
Hamiltonians like $H_{\bf HF}$ and not as accidents caused by unfortunately chosen parameters. All this suggests that a certain amount of disorder 
has to be present to account for the observed {\it stationarity} of the MH-band in LSCO, whereas for Na-CCOC possibly the weak-disorder scenario is the 
more appropriate one. However, because an inclination to chemical potential pinning does exist (Fig.\ref{fig:mc_nw}), disorder strength may not be overly large 
to induce the effects described here.      
Note, however, that at this point it is difficult to decide whether such solutions simply result from the original mean-field approximation or represent 
genuine low-energy states for doped MH insulators and further investigations may be needed. 
\section{Spectral Functions in the Presence of Competing States}
The BdG equations were solved on lattices of the size $N$=40$\times$40, which allows for a detailed comparison with experimental data. 
Although larger systems could in principle be used, the great number of iterations necessary to achieve self-consistency at temperatures 
not-too-far from clean-case critical temperatures renders such systems practically useless. We have also observed 
that as the lattice is increased the convergence at high temperatures suffers and additional iterations are required to arrive at a self-consistent solution. 
As in the previous effort, charge-depleted ``plaquettes'' were distributed on the lattice in a random fashion. Those plaquettes were allowed 
to overlap, but care was taken to ensure that the area covered (i.e. the relative amount of sites with reduced chemical potential) grew linearly with 
doping. Since the overlap probability increases with the number of plaquettes already present, their number has to grow supralinearily with the hole concentration. 
In this fashion, however, truly random configurations can be produced, and the corresponding self-consistent equations solved. 

One such example can be seen in Fig.\ref{fig:definelabel}(a)-(c), where $\Delta_{\bf i}$ and $m_{\bf i}$ are shown for 
hole concentration $x$=0.15 at three different temperatures, starting from $T$$\sim$0 (Fig.\ref{fig:definelabel}(a), (a1)) to a finite temperature $T$$<$$T^{SC}_c$ 
((b), (b1)) (see Section V), and finally at $T$$\approx$$T^{SC}_c$ ((c), (c1)). Given our values for $n_{\rm SC}$=0.75 and $n_{\rm AF}$=1, 
about 40$\%$ of the total area are insulating. 
Dark color translates to large OP amplitude, chosen in a way so that the darkest possible ones coincide with the corresponding amplitude 
in the clean case. As far as the SC OP is concerned, an increase in temperature at first transforms sites at the boundaries into metallic ones, 
but leaves the OP strength inside larger clusters unaffected (Fig.\ref{fig:definelabel}(b)). Eventually, this leads to a breakup of the original SC ``backbone'' into a set 
of single, isolated islands (Fig.\ref{fig:definelabel}(c)). Those small clusters, however, are capable of surviving to surprisingly large temperatures, in this particular 
case up to $T$$\sim$0.28 (3/4$T^{SC}_c$). The local AF order, on the other hand, appears more stable as it persists almost up to $T^{AF}_c$ with only a small 
loss of overall volume (Fig.\ref{fig:definelabel}(c1)), and thus there exists an extended temperature range where metallic and AF regions coexist. 
This observation can be made regardless of the overall amount of insulating clusters. The phase diagram that can be derived from data like 
those will be unveiled in Section V.  

\begin{figure} 
\includegraphics[width=8.0cm,clip]{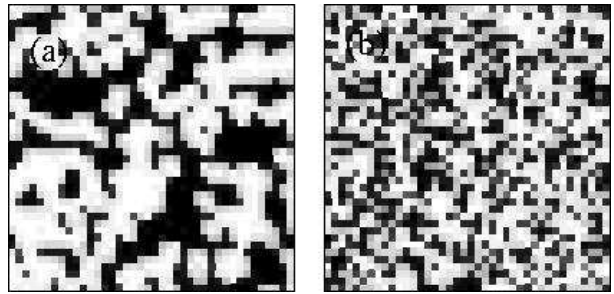}
\caption{Influence of the plaquette geometry on the local AF OP distribution. (a) corresponds to the procedure of using extended plaquettes, mainly adopted in this work, 
whereas in (b) the size of the plaquettes is 1, leading to a more ``fractal'' shape of the resulting clusters. Particle density $\langle$$n$$\rangle$=0.85 
and $T$$\simeq$0 in both cases.} 
\label{fig:plaquette}
\end{figure}

\begin{figure}
\includegraphics[width=7.0cm,clip]{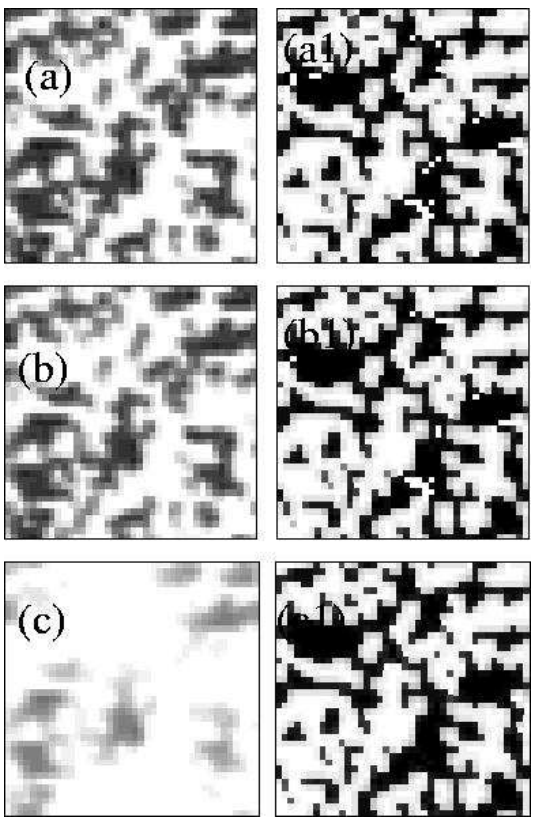}
\caption{\label{fig:definelabel} Order parameters $\Delta_{\bf i}$ (left column) and $m_{\bf i}$ (right column) at different 
temperatures $T$$\sim$0 ((a), (a1)), $T$=0.10 ((b), (b1)), and $T$=0.25 ((c),(c1)) at overall density $\langle$$n$$\rangle$=0.85.}
\end{figure}

The choice of plaquettes, i.e., their size, does have certain effects on the outcome, and this is demonstrated in Figs.\ref{fig:plaquette}(a), (b), which shows 
$m_{\bf i}$ for (a) the procedure used throughout the paper, and (b) where a large number of very small (i.e., of size 1) ``plaquettes'' were used \cite{comment_plaq}.  
In the latter case, the clusters appear much more irregular and fractal, as expected. Although the resulting spectral functions are hard to distinguish, 
and thus our results not obviously affected by any particular choice, it does influence the OP distribution function as we will discuss in a later section.
 
\begin{figure}
\includegraphics[width=8.5cm,clip]{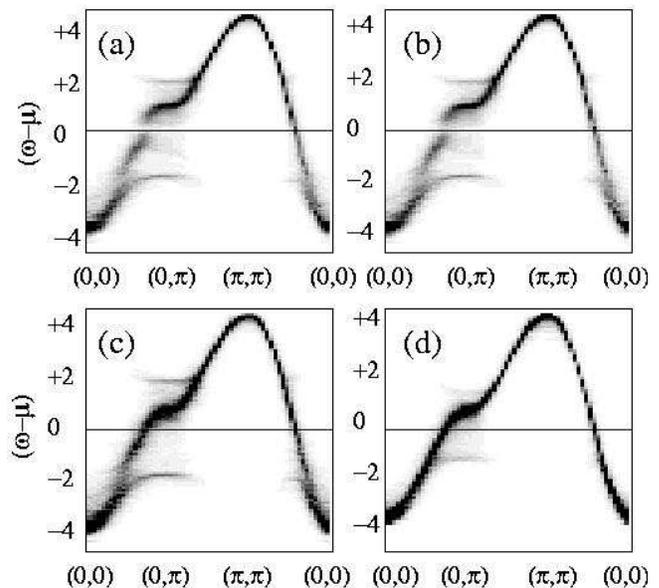}
\caption{\label{fig:fs_arc} $A$(${\bf k}$, $\omega$) at density $\langle$$n$$\rangle$=0.85 at 4 different temperatures: $T$$\sim$0 (a), $T$=0.10 (b), 
$T$=0.30 (c), and (d) $T$=0.80. Even though a FS is established for ${\bf k}$$\sim$($\pi$,0) at intermediate temperatures, the SDW branch survives up to the largest 
temperatures and is responsible for the depletion of the low-energy density of states.}
\end{figure}

The spectral functions related to mixed AF/SC states such as in Fig.\ref{fig:definelabel} are shown in Fig.\ref{fig:fs_arc}; 
those for the two ``clean'' cases have already been presented in the previous paper \cite{Mayr_1} and are therefore not displayed here. 
There, it was also demonstrated that the spectral functions in the mixed state can effectively be obtained by superimposing the spectral 
functions of the respective clean phases on top of each other, provided the two signals are weighted by the relative amount of either phase.
Thus, at the lowest temperatures considered, both the two-peak structure 
at ($\pi$,0) and (although it can be barely seen here) at ($\pi$/2, $\pi$/2) is visible as is the presence of a small gap at $E_{\bf F}$ 
in the anti-nodal direction (Fig.\ref{fig:fs_arc}(a)). The spectra themselves are ``smeared out'' and poorly defined, especially in the anti-nodal direction, 
similar to what is observed in ARPES. The AF branch is found at $\omega$$\simeq$-2, consistent with Fig.\ref{fig:dos}, whereas the maximum of the low-energy intensity 
is at $\omega$$\sim$-0.5, but the corresponding distribution is very broad and intensity is found almost up to $E_{\rm F}$ \cite{comment_den}.  
As $T$ is increased and SC states become metallic ones (see Fig.\ref{fig:definelabel}), one observes the accumulation of spectral weight 
right at $E_{\rm F}$ (Fig.\ref{fig:fs_arc}(c)). This takes place in the temperature window somewhat (but not too much) below $T^{SC}_c$, 
and the FS crossing becomes much better defined for higher temperatures (Fig.\ref{fig:fs_arc}(d)). Nevertheless, remnants of the insulating regime 
still survive at those temperatures, and the two-peak structure in A(${\bf k}$, $\omega$) is preserved up to almost $T^{AF}_c$. 
Note, however, that the SC phase is replaced by a metallic one and not by the competing AF state. The latter scenario is certainly also possible, 
particularly in the clean case, 
and if a first-order phase transition is separating both phases. In such a case, the signal close to $E_{\rm F}$ would 
actually vanish at elevated temperatures, and the only remaining gap would be the one corresponding to the LHB. Unfortunately, ARPES as of now, cannot tell us 
what is happening since measurements in this temperature range are currently not possible \cite{comment_yoshida}. 
As long as disorder is important, however, we think that the presented scenario is the correct one. 
The position of the LHB in any case is strongly influenced by the temperature as well, quite unlike what is observed if the carrier-density is changed. 
It shifts to lower binding energy, and weakens at the same time.  

At this point we want to emphasize that the transition between different regimes is a very ``soft'' or gradual one, and particularly the loss of long-range SC 
order is not coinciding with any dramatic changes in the spectral functions, unlike in the case of the common BCS transition, which entails the sudden 
emergence of Bogoliubov quasiparticles. Fig.\ref{fig:fs_arc} testifies to that. With the demise of superconductivity one should, of course, 
expect the emergence of a complete FS somewhere 
below $T^{SC}_{\rm c}$, but this is not what is occurring here, at least in the weakly doped regime: since only a small volume percentage is affected by the 
SC/metal transition, and with most of the spectral weight residing in the AF branch, the changes in the spectral functions are minuscule and 
they do not resemble those of a true qp. Clearly, the properties of the sample are dominated by the presence of insulating regions. 
All this is quite typical for systems close to a percolation threshold, where a small change in volume fractions (which are effectively 
measured by ARPES) can induce a drastic change in transport properties. 
\subsection{Proximity effects in inhomogeneous states}
There has been considerable progress in the quality of data provided by PE experiments since the first results have become available. Whereas originally 
ARPES results only revealed the large excitation gap of about $\omega_{\rm E}$$\simeq$0.2 eV ($x$=0.03) in the anti-nodal direction, later results have shown that spectral 
weight can in fact be found for {\it all} energy values between $\omega_{\rm E}$ and $E_{\rm F}$\cite{Yoshida_2}. Nevertheless, it was confirmed that the main intensity 
indeed resides at $\omega_{\rm E}$; this main intensity peak moves towards $E_{\rm F}$ at subsequent doping, but remains poorly defined, and becomes qp-like 
only close to optimal doping. 
\begin{figure}
\includegraphics[width=8.0cm,clip]{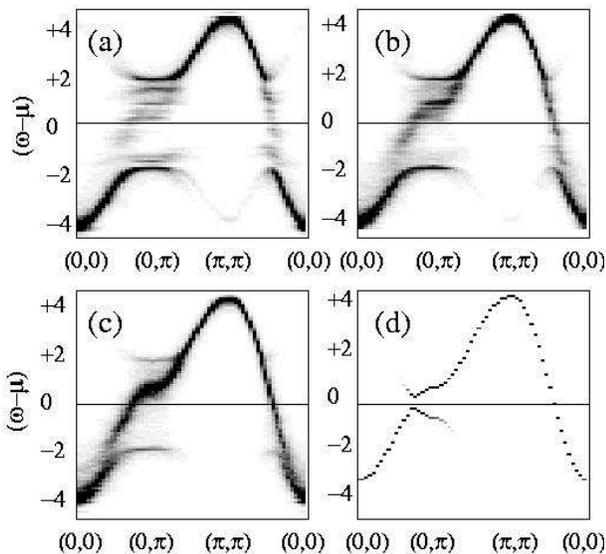}
\caption{\label{fig:v06}The spectral functions for a mixed AF-metallic model, with $U$=5 and $V$=0.0 and densities (a) \nn =0.95, and (b) \nn =0.90. 
Even though the metallic clusters are ungapped courtesy of $H_{\bf HF}$, there is yet a finite excitation gap in 
the anti-nodal direction close to half-filling ((a)), caused by proximity effects. The same data as in (a), but for $V$=0.6, are plotted in (c) 
and (d) is the spectral function for the homogeneous superconductor with $V$=0.6 for \nn=0.75. The shrinking of the gap close to ($\pi$, 0) between (c), (d) is obvious.}
\end{figure} 
Fig.\ref{fig:dos} does not fully reproduce this behavior, since it suggests that the gap is actually {\it increasing} with doping rather than decreasing. 
Certainly the initial large value of $\omega_{\rm E}$ can be described in terms of a very strong attraction or an additional, 
but hitherto hidden, OP such as DDW, by making $V$ a function of $x$; 
nevertheless, in the scenario where doping is achieved by replacing hole-poor sites with hole-rich ones, it is mysterious as to why 
those parameters are changing even though the (local) density remains constant. Here we attempt a different explanation for the shrinking of the gap, 
and one that does not rely on additional assumptions beyond those of the existence of a mixed-state. 


To do so we first remark that the choice of parameters in Ref.\onlinecite{Mayr_1} (as well as in this work) has been somewhat unfortunate as the distinction 
between the AF and the SC energy scales that is present in the cuprates is lost to some degree, for $V$=$U$/5 is unnaturally large, at least as far as LSCO is concerned. 
The self-consistent solution of Eq.(\ref{eq:hamfermi0}) tells us that in a mixed-phase sample due to proximity effects there are always sites or regions that carry 
{\it induced} OP amplitudes, namely those at or close to phase boundaries (Fig.\ref{fig:definelabel}). Therefore, sites that are supposed to be solely SC, 
due to the local term in $H_{\bf HF}$, may have a finite magnetization as well and vice versa. 
To better understand the relevance of such effects, 
$H_{\bf HF}$ was solved again, but this time with $V$=0 in order to disentangle the AF from the SC OP. This is important because in actual materials such 
proximity effects, due to the large energy associated with AF ordering, may overshadow the influence of the SC term and actually determine the observed gap. 
In contrast, for the particular choice of $V$/$U$ made here as well as in the previous work, the ``clean'' SC gap $\Delta_{\rm S{\bf i}}$=$V\cdot\Delta_{\bf i}$ 
is about as large as the magnetic gap $\Delta_{\rm m}$=$\frac{U}{2}$$m_{\bf i}$ on sites ${\bf i}$ located next to an AF cluster.
Therefore, we reduce $V$ in an effort to correctly describe the properties of underdoped cuprates, starting out with the simplest case, $V$=0. 
The spectral functions now emerging are shown in Figs.\ref{fig:v06}(a), (b). 
Even though the low-energy intensity is not very significant, one can conclude that most of the low-energy weight in the anti-nodal direction 
is concentrated at binding energy $\omega$$\sim$-0.4 for $x$=0.05 and appears to have a finite gap, although the corresponding 
interaction $V$ has been set to zero! An inspection shows that this finite gap is caused by proximity effects, which dominate for small doping levels. 
As doping progresses, this intensity travels further towards $E_{\rm F}$, 
and actually starts to cross $E_{\rm F}$ for 0.05$<$$x$$<$0.10. Although the signals are blurred due to the inhomogeneous nature of the sample, the main 
issue, the existence (or enhancement) of excitation gaps close to $\langle$$n$$\rangle$=1 is clearly demonstrated in Fig.\ref{fig:v06}(a). 
We also want to reiterate here that the same ``blurriness'' in the spectral intensities has been observed for samples with intermediate ($x$=0.07) hole 
concentration\cite{Yoshida_2}.

\begin{figure}
\includegraphics[width=8.0cm,clip]{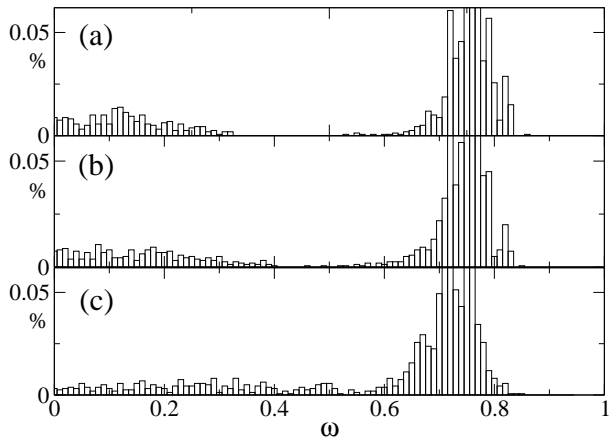}
\caption{$N$($m$) as a function of energy for the density $\langle$$n$$\rangle$=0.95 (i.e., 20$\%$ non-AF sites). 
The main peak at $\omega_{\rm m}$$\simeq$0.75 pertains to the homogeneous AF phase, whereas the low-energy values stem from boundary sites. 
(a) is the distribution for the 12-site plaquette as mostly used in 
this work, (b) is the 6-site plaquette, and (c) the 1-site plaquette. The samples become more fractal when going from (a) to (c), and this 
is reflected in the very broad OP distribution function.}
\label{fig:op_dist}
\end{figure}

Furthermore, similar conclusions can be achieved by directly measuring the AF OP distribution, $N$($m$), which is depicted in Fig.\ref{fig:op_dist} 
for $x$=0.05, and which serves to substantiate and clarify our arguments. For such a small number of holes, $N$($m$) has the expected 
peak at $\omega_{\rm m}$$\sim$0.75 (thus, $\Delta_{\rm m}$$\approx$2, see Fig.\ref{fig:dos}) in parallel with the one found for half-filled systems 
and stemming from sites that are located deep in AF territory, {\it plus} low-energy contributions below $\omega$$\approx$0.3. 
They stem from $U$=0 sites in Eq.(\ref{eq:hamfermi0}), that suffer from proximity effects.   
For $x$=0.05 there are only few charge-depleted, uncorrelated clusters, and many such sites are in close proximity to hole-poor ones. Therefore, the low-energy 
distribution is peaked at a comparatively large value, $\omega_{\rm le}$$\sim$0.12, although the distribution itself is somewhat broad (see Fig.\ref{fig:op_dist}(a)). 
This distribution, however, depends to quite some degree on the specific shape of plaquettes chosen, which is demonstrated in Figs.\ref{fig:op_dist}(b), (c). 
The latter one, corresponding 
to a OP distribution such as depicted in Fig.\ref{fig:plaquette}(b) shows that for a sufficiently fractal configuration, the AF OP can assume {\it any} value between 
zero and $\omega_{\rm m}$, whereas the former one resides between the two extremal cases considered here, and has a distribution peak at $\omega_{\rm le}$$\sim$0.2. 
Although this ratio is closer to the actual ratio 1/3 (0.2eV vs. 0.6eV), our arguments are not affected by any particular choice, and we will mostly stick to 
our original geometry.      

\begin{figure}
\includegraphics[width=8.0cm,clip]{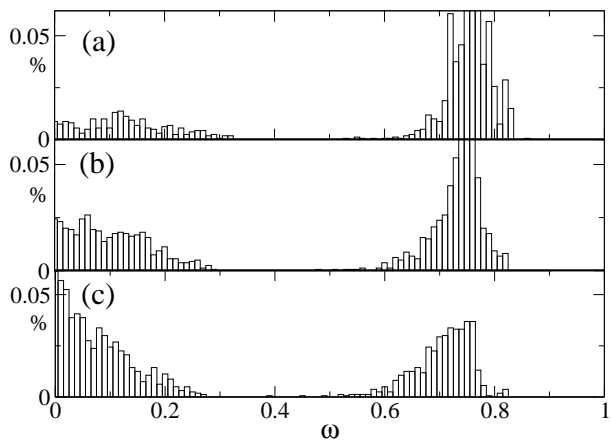}
\caption{$N$($m$) at low temperatures as a function of energy for varying densities: 
(a) $\langle$$n$$\rangle$=0.95, (b) $\langle$$n$$\rangle$=0.90, and (c) $\langle$$n$$\rangle$=0.85 in the mixed-phase regime. 
$V$=0 for non-AF sites.}
\label{fig:op_dist1}
\end{figure}

\begin{figure}
\includegraphics[width=8.0cm,clip]{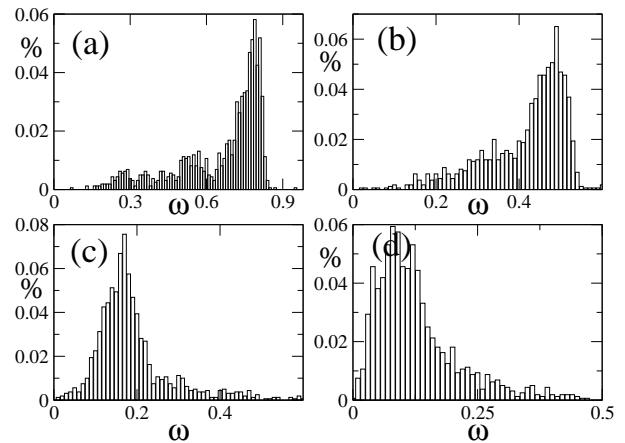}
\caption{Relative distribution of the total gap $\Delta_{\rm eff}$ for the density $\langle$$n$$\rangle$=0.80 (i.e. 80$\%$ SC) for 
$V$=1 (a), $V$=0.80 (b), $V$=0.60 (c), and $V$=0.60, but a more fractal configuration (d). The best agreement with STM data is found for (c).}
\label{fig:gap_dist}
\end{figure}

In any case, adding holes changes $N$($m$) considerably, as it becomes dominated by non-interacting sites and the peak of the low-energy part of $N$($m$) is 
shifted towards $\omega_{\rm le}$=0. This is demonstrated in Fig.\ref{fig:op_dist1}, which shows the $N$($m$) at different densities as one 
progresses towards $x_{\rm c}$. Whereas initially (Fig.\ref{fig:op_dist1}(a), same as Fig.\ref{fig:op_dist}(a)) the low-energy part is dominated by 
finite energy contributions, 
subsequently the main peak in the distributions shifts to smaller energies (Fig.\ref{fig:op_dist1}(b)). Thus, at the onset of doping, the majority of states 
has a larger excitation gap than predicted by the value of $V$ (which is actually zero here), and it is those states that have the largest weight and, 
therefore, intensity in PE; 
still, there are a few states with a very small (or zero) $m_{\bf i}$, and they are responsible for the finite spectral weight at $E_{\rm F}$ (along the BZ diagonal).
As doping progresses, less and less metallic sites will be influenced by neighboring insulating ones, and eventually the vast majority of sites will have only a 
very small or even zero magnetization (Fig.\ref{fig:op_dist1}(c)), causing the spectral intensity peak to move towards $E_{\rm F}$. A simple calculation confirms this 
picture: the average magnetic gap $\bar \Delta_{\rm m}$ (counting only the {\it low-energy} contributions) 
changes from $\bar \Delta_{\rm m}$=0.33 ($\langle$$n$$\rangle$=0.95) to $\bar \Delta_{\rm m}$=0.26 ($\langle$$n$$\rangle$=0.90), 
to $\bar \Delta_{\rm m}$=0.20 ($\langle$$n$$\rangle$=0.85), and finally to $\bar \Delta_{\rm m}$=0.11 ($\langle$$n$$\rangle$=0.80). 
These results refer to Figs.\ref{fig:op_dist1}(a)-(c) but similar ones were obtained for other configurations as well and demonstrate 
that any low-energy gap in mixed insulating/metallic systems decreases {\it continuously} upon replacing insulating with metallic sites. 
An identical behavior was observed in STM as well\cite{McElroy_1}.     
Choosing a finite $V$ will not change these results aside from adding a small additional gap (if $V$ is small compared to $U$) 
and preventing $A$(${\bf k}$, $\omega$) from actually developing a FS crossing. For any given site, the local effective gap (at least in the mean-field approximation) 
will then be $\Delta_{{\rm eff}, {\bf i}}$ = $\sqrt{\Delta^2_{\rm S {\bf i}} + \Delta^2_{{\rm m}{\bf i}}}$, i.e., it is a combination of both AF and SC gaps. 
The relative distribution of $\Delta_{\rm eff}$ ($\langle$$n$$\rangle$=0.80), $N$($\Delta_{\rm eff}$), is shown in Fig.\ref{fig:gap_dist} for different values of $V$. 
The results for $V$=1, the value chosen in Ref.\onlinecite{Mayr_1}, can be found in Fig.\ref{fig:gap_dist}(a), and most importantly, 
$N$($\Delta_{\rm eff}$) is peaked at the comparatively large energy $\omega$$\sim$0.75 (compare, e.g., with Fig.\ref{fig:dos}), and thus $\Delta_{\rm S}$ 
dwarfs residual AF gaps ($\Delta_{\rm m}$$\sim$0.1). In this case, the average gap $\bar \Delta_{\rm eff}$ actually {\it increases} with hole-doping, 
as relatively small proximity-induced AF gaps are replaced by large SC ones. Then, a much better choice to understand PE results in LSCO is 
found in $V$=0.6 (Fig.\ref{fig:gap_dist}(c)), where both $\Delta_{\rm S {\bf i}}$, $\Delta_{\rm m{\bf i}}$ (in hole-rich areas) 
are of the same order, and $N$($\Delta_{\rm eff}$) is peaked slightly 
below $\omega$=0.2. In this case, $\bar \Delta_{\rm eff}$ decreases with the addition of holes, from the original value ($x$=0.05) $\bar \Delta_{\rm eff}$$\sim$0.33 
(superconductivity is almost negligible there) to, eventually, $\bar \Delta_{\rm eff}$$\approx$0.2, where the gap is mostly defined by $V$. In other words, we argue that 
the observed large effective gap at small hole concentrations is caused by intermixing antiferromagnetism via proximity effects, which is then gradually replaced by 
the pure SC gap as doping progresses. This provides in our view the simplest and most natural explanation for the doping dependence of the LSCO low-energy gap. 
This picture agrees also with the recent notion that this gap is composed of a low-energy contribution following $d$-wave behavior, and a larger one, 
leading to an overall deviation from the simple $\cos k_x$-$\cos k_y$ form\cite{Ronning_1}.
Figs.\ref{fig:v06}(c),(d), which show the anti-nodal gap for the same configuration as before, but with $V$=0.6, 
changes from $\omega_{\rm E}$$\sim$0.5 to  $\omega_{\rm E}$$\sim$0.2 between \nn=0.95 and \nn=0.75 confirms this once more as does $N$($\omega$), where the 
small original peak (Fig.\ref{fig:dos}) grows {\it and} evolves towards $E_{\rm F}$ (i.e., $\omega$$\sim$0.2) with doping. Yet, we stress that proximity effects are 
important since without them the peak associated with superconductivity simply would remain in its position.
The critical temperature for $V$=0.6 is $T_c$$\sim$0.14, and thus the respective clean case AF/SC critical temperature ratio $\approx$1/7, reasonably 
close to the actual LSCO ratio 1/10. Although this leads us to believe that an even more faithful representation of ARPES results can be achieved
by choosing a smaller value of $V$, the resulting difficulties in the self-consistent scheme prevent us from doing so.  
Already, however, $N$($\Delta_{\rm eff}$) in Fig.\ref{fig:gap_dist}(c) strongly resemble the one found via low-bias STM in Bi-2212 \cite{Pan_1}: 
its shape is almost Gaussian, with only a minor tail, quite unlike Fig.\ref{fig:gap_dist}(a), (b), (d). We also mention here that in Ref.\onlinecite{Pan_1} 
the gap distribution is mirrored by a similar charge distribution, as is the case in our model. A logical next step is the calculation of $N$({\bf i}, $\omega$) 
and a detailed comparison with STM data. Unfortunately, this calculation suffers from serious 
finite-size effects\cite{Atkinson_2}, and important questions such as whether sites associated with very small (large) $m_{\bf i}$ are associated with a 
comparatively prominent (small) coherence peak cannot satisfactorily be resolved. For this reason, we forgo a discussion of $N$({\bf i}, $\omega$) for the time 
being.   

Our goal here was to demonstrate that it is necessary to carefully chose the parameters to fully reproduce the LSCO PE data. 
Proximity effects ought to be ubiquitous in mixed-state phases, and it has now been shown that their effects have measurable consequences. 
Neutron diffraction experiments by Lake {\it et al.}\cite{Lake_1} showed that the AF response in underdoped LSCO {\it increases} upon the 
application of a magnetic field, which supresses superconductivity, a result which may be consistent with the existence of SC/AF coexisting 
regions in inhomogeneous materials, where the same behavior can be expected.

\subsection{The Fermi surface}
\begin{figure}
\includegraphics[width=8cm,clip]{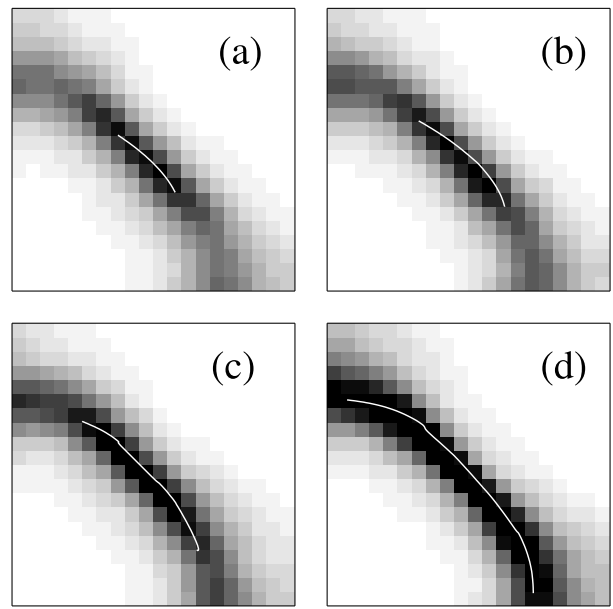}
\caption{\label{fig:arpes_bdg} Spectral weight at the Fermi level at different temperatures for $H_{\bf HF}$: $T$$\sim$0 (a), $T$=0.20 (b), $T$=0.50 (c), and $T$=0.80 (d).
The original ``arc'' (indicated by white lines) expands as the temperature is increased, and becomes a full FS for temperatures much larger than $T^{SC}_c$. 
Even then, because of residual insulating clusters, the FS in the anti-nodal directions is not particularly well defined.}
\end{figure}

The corresponding FS can be determined by measuring the spectral weight at $E_{\rm F}$ (integrated over a small energy window, 
$\Delta$$\omega$$\sim$$0.2$-$0.3$), and it is shown in Figs.\ref{fig:arpes_bdg}(a)-(d) as it evolves with temperature (\nn= 0.90). 
Close to $T$$\sim$0, significant weight (and a qp-like peak) is found only in the neighborhood close to the BZ diagonal, as already demonstrated 
in the previous paper \cite{Mayr_1}. The resulting FS ``arc'' widens as the temperature is raised, indicating enhanced metallicity, until it becomes a 
full (electron-like) FS at about $T$=0.80 (Fig.\ref{fig:arpes_bdg}(d)). 
Nevertheless, even for the temperatures distinctively above $T^{SC}_c$, does the FS not fully resemble the one of a conventional metal. In fact, even 
for $T$$>$2$T^{SC}_c$ does the FS appear ``blurred'' at momenta ($0,\pi$), ($\pi$,0), and only above $\sim$$T^{AF}_c$ is a regular, sharply-defined FS recovered. 
Thus, if the PG is caused by a remnant insulating phase, it must at the same time be characterized by a poorly defined FS at the anti-nodal points. 
Similar results are found for other densities as well. In particular, this has important implications for the SC transition for $x$$>$0.10 (see Fig.\ref{fig:pd}); 
in a conventional superconductor, this transition is accompanied by the restoration of the underlying FS in the anti-nodal direction, something that does not happen 
in the case regarded here. 
Therefore, the change from the SC phase into a ``metallic'' one will appear as much more gradual, and less abrupt than a conventional SC/metall transition.  

In the mixed regime, the {\it length} of the FS arc is to a large degree {\it independent} of the hole density, whereas the {\it spectral intensity} at $E_{\rm F}$ 
grows continuously. This observation, made in PE experiments and later reproduced in calculations based on the AF/SC mixed-state 
picture \cite{Yoshida_1, Mayr_1}, is naturally interpreted as a consequence of the increased relative weight of the SC phase, 
and subsequently, of the importance of zero-energy excitations, within the sample upon doping. 
Those observations are somewhat {\it reversed} when the temperature dependence of the FS is considered: 
as the arc is spreading out with enhanced $T$ ($T$$<$0.8 for the parameters considered here), the {\it intensity} (along the BZ diagonal) stays the 
same (within 15$\%$); this so happens because SC regions are turned into metallic ones, and the states most affected by that are the ones that are 
close to the BZ diagonal, where $\Delta$ is smaller. 
States along the BZ diagonal are not affected since they are already metallic. Nevertheless, the spectral intensity along the anti-nodal direction, 
which is independent of doping to a large degree, shows a distinct increase with 
temperature within the mixed-state scenario as some marginally AF sites are turned into metallic ones. Note, that it is only important in this context 
that the SC/AF energy scale are sufficiently separated so 
there exists a broad temperature range without SC order, but an intact AF phase. This certainly applies to the cuprates, and therefore, these predictions 
are verifiable within PE experiments as long as the necessary temperatures can be achieved.   


Furthermore, the scenario presented here naturally explains the behavior of the (integrated) spectral weight at $E_{\rm F}$. It was shown 
in Ref.\onlinecite{Yoshida_1} (Fig.4 therein) that this quantity (as well as the qp-density derived from Hall effect measurements) grows continuously with doping, 
which naturally reflects the increased amount of metallic areas in the sample, as already remarked in Ref.\onlinecite{Mayr_1}. Yet, the rate of increase is 
comparatively small for $x$$<$0.1, but increases strongly thereafter, whereas this rate is constant following Ref.\onlinecite{Mayr_1}. 
Proximity effects provide us with an explanation: close to the parent insulator, many nominally ``free'' sites are still to some degree AF ordered, 
and the metallic volume is smaller than 
\begin{table}[hbtp]
\hspace*{-0.2cm}
\caption{Integrated spectral intensity near the Fermi level as a function of the electron density, $\langle$$n$$\rangle$.}        
\centering
\begin{tabular}{lc|cccccccccc}
    $\langle$$n$$\rangle$  & &  $I_1$  & & & $I_2$ & & & $I_3$ \\ \hline
   0.95 & & 13.81   & & & 15.48 & & & 27.95   \\        
   0.90 & & 36.55   & & & 40.52 & & & 72.98   \\        
   0.85 & & 54.68   & & & 61.12 & & & 107.1   \\        
   0.80 & & 74.50   & & & 82.57 & & & 142.05   \\ \hline\hline 
\end{tabular}
\label{table_1}
\vspace*{-0.1cm}
\end{table}
one would expect according to the doping level. To make up for the ``lost'' sites - which has to happen eventually at optimal doping - their number has to 
increase at a faster rate at higher doping levels. To underline this point we have integrated the near-$E_{\rm F}$ spectral weight 
(within a window of $\Delta$$\omega$=0.3eV), again for the case of $V$=0, and have tabled those values, $I_{1-3}$, in Table I. $I_1$ ($I_2$) 
refer to different ${\bf k}$-space 
windows (along the BZ diagonal), whereas $I_3$ refers to data obtained by using $\Delta$$\omega$=0.4, but the same BZ cut as in $I_1$.
Irrespective of how the window is defined, we find that the rate of increase is small for $x$$<$0.05, but considerably larger afterwards. 
It is quite impressive that even such details of PE data can be reliably reproduced within the simple mixed-state picture.

Still, even at low temperatures, the high-intensity spectral weight together with the low-intensity signals 
in the anti-nodal direction (Fig.\ref{fig:arpes_bdg}(a)) evokes memories of a complete FS, as one would expect it for a non-interacting 
system of $\sim$$n_{\rm SC}$ electrons. Experimentally, it was demonstrated that at least in the case of LSCO a FS constructed in this way approximately 
fulfilles Luttinger's theorem \cite{Yoshida_2}, and a similar result was found theoretically as well \cite{Mayr_1}. 
This interesting result suggests that the underdoped phase is actually closely linked with the weakly-interacting metal, and is, for example, 
not separated from it by any FS transition. Our results here imply that the FS that ought to emerge as the temperature is raised 
is the one of the underlying, hole-rich metallic phase, which in itself is independent of the overall doping level, that is, 
it should be the same FS independent of doping. ARPES data at elevated temperatures could confirm this picture.  


\section{Phase diagram}
Based on the considerations presented above it is tempting to develop a generic phase diagram of $H_{\rm HF}$ in the $T$-$x$-plane, across the whole doping range. 
This is relatively straightforward in the homogeneous case, where the condition $\Delta$($T_c$)$\equiv$0 determines the critical lines. In the inhomogeneous scenario 
presented here, a more careful approach needs to be taken in order to distinguish between {\it global} and {\it local} phenomena, both of which characterize the sample as 
discussed above. In the mean-field scheme employed here, however, a true distinction between an actual critical (global) temperature $T_c$ 
and a ordering temperature $T^{\star}$ is difficult to achieve. This is not unlike the problem encountered in  
the strong-coupling limit, where, in the context of superconductivity, Cooper pairs form at some large $T^\star$, but condense in a Kosterlitz-Thouless 
transition only at a $T_c$ much lower. Nevertheless, the ideas presented here have more in common with those discussed, e.g., in diluted magnetic semiconductors, 
where, in the limit of small impurity doping, one may observe the formation of isolated, locally ordered clusters at a large temperature $T^{\star}$; 
as the temperature is lowered the associated 
wavefunctions expand and start to overlap at $T_c$ in a percolative transition \cite{Mayr_3}.
We assume that this scenario gives the best description of the SC transition in the underdoped regime. Therefore, we adopt the viewpoint that 
a phase transition in the inhomogeneous regime occurs once a certain percentage of lattice sites assumes a finite-value of the OP, either AF or SC, and 
chose here as the critical 
percentage $p_{\rm c}$=60$\%$. This value is motivated by extensive MC calculations for site-diluted 2D Heisenberg models that have shown that the percolation 
transition (at $T$=0) is essentially a {\it classical} one, i.e., long-range order (LRO) is lost once $\sim$40$\%$ of the spin-carrying sites are removed, 
in agreement with (classical) bond-diluted random resistor networks \cite{Sandvik_1, Kirkpatrick_1}. Since the SC phase transitions are in the 
same (XY) universality class, we adopt this particular value for that case as well. Note that because of the previously discussed proximity 
effects the relative amount 
of sites carrying a finite OP is actually larger than the number of sites with a finite value of $U$ or $V$ in $H_{\bf HF}$; this is the main effect 
of the hopping terms in the Hamiltonian when 
compared to the localized spin models employed in MC simulations. Even if the actual $p_{\rm c}$ differs from the chosen value - and this can only be 
verified by a more thorough investigation into site-or bond-diluted superconductors, presumably with more refined MC simulations - 
the shape of the $T_c$($x$) curves is not supposed to be extraordinarily affected by any particular choice of $p_{\rm c}$. 

\begin{figure}
\includegraphics[width=8.5cm,clip]{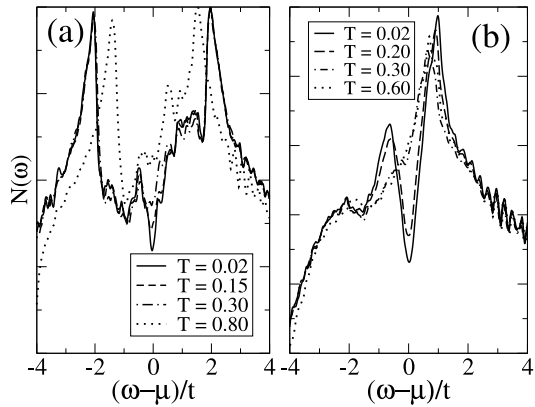}
\caption{\label{fig:dos_t} Temperature-dependent $N$($\omega$) for two different doping levels \nn=0.90 (a) and \nn =0.80 (b). Whereas in (a) 
the gap-like feature survives almost up to the clean transition temperature $T^{AF}_c$$\sim$1, it is barely notifiable in (b) at temperatures just above $T_c$.}
\end{figure}

Unfortunately, it turns out that even for a strongly doped system the AF OP in isolated pockets survives almost up to $T^{AF}_c$, 
probably an unfortunate consequence of the approximation involved. Therefore, we have in addition calculated the approximate temperature 
upon where the associated gap in $N$($\omega$) closes and defined this value as $T^\star$. 
This procedure should not be regarded as particularly unusual, since (a) it is a common way to 
determine $T^{\star}$ in underdoped cuprates anyway, and (b) because $T^{\star}$ in itself is not a strictly defined quantity. Rather, it is a temperature much larger than 
$T_c$ below which certain excitations are suppressed, and the actual values cited in the literature vary depending on the experimental quantity investigated, 
or even allow for the definition of multiple PG temperatures. 

\begin{figure}
\includegraphics[width=8.5cm,clip]{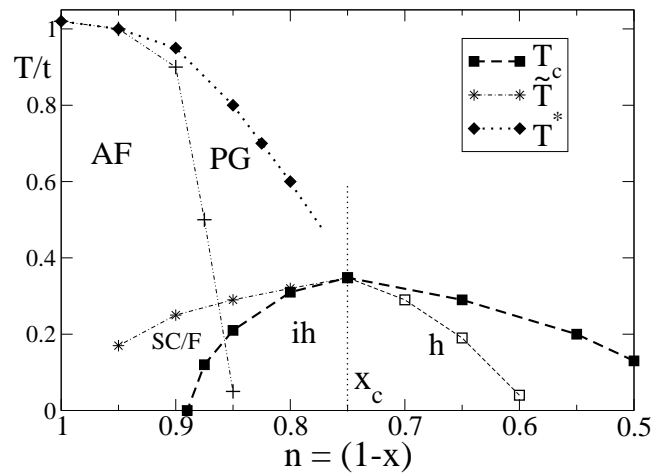}
\caption{\label{fig:pd} Phase diagram for the cuprates, derived from Eq.(\ref{eq:hamfermi0}), where hole doping leads to a coexistence of SC clusters 
(density $n_{\rm SC}$$\sim$1-$x_{\rm c}$) with the parent Mott insulator. The line of optimal doping $x_{\rm c}$ separates two fundamentally different regimes, 
the homogeneous (h) overdoped phase, where rather common BCS physics can be expected, and the inhomogenous (ih) underdoped regime, 
which consists of the AF/SC mixture and whose SC transition is of percolative nature. $T_c$ determined as described in the text, lines serve as guide to the eyes.
Local SC clusters survive up to $\tilde T$, which forms the upper bound of a SC fluctuating (SC/F) regime, 
and $T^\star$ denotes the PG temperature as derived from the density of states. The existence of the comparatively large 
gap for small doping levels allows for the definition of an additional characteristic temperature, if so desired. This temperature, however, is poorly defined 
and thus omitted here. We have also indicated the boundaries of the AF phase. Assuming that $V$ is in some sense derived from spin fluctuations, 
the $T_c$($n$) curve in the overdoped regime is replaced by the thin broken line, and similarities to the conventional phase diagram become more pronounced.}
\end{figure}

The resulting phase diagram (for the set of parameters employed throughout this work) is shown in Fig.\ref{fig:pd}. It is characterized by a 
multitude of different relevant temperatures in the underdoped regime, such as 
the critical (SC) temperature $T_c$ (thick broken line), a temperature $\tilde T$ up to where {\it local} superconductivity survives (indicated by stars in Fig.\ref{fig:pd}), 
and a PG temperature $T^\star$, determined 
by the gap in $N$($\omega$). Most striking is the similarity to the standard cuprate phase diagram, with a PG phase and the SC dome and it is noteworthy 
that this has been achieved within such a simple scheme and without any further assumptions. In the framework of the mixed-phase state as discussed here, 
the central role is assumed by the level of optimal doping, which separates the inhomogeneous and homogeneous parts of the phase diagram from each other. 
For doping levels smaller than 
$x_{\rm c}$, doping is tantamount to replacing AF-ordered sites with SC ones, whereas the density {\it within} those SC clusters remains unchanged (at least in a 
first approximation). 
For $x$$>$$x_{\rm c}$, on the other hand, further doping actually does lead to a decrease of the charge density across the whole sample, 
and subsequently to a reduction in $T_c$. The existence of a maximum in $T_c$ is therefore a natural consequence in such a scenario. It needs to be noted, however, that 
according to Fig.\ref{fig:pd} the reduction in $T_c$ is much smaller in the overdoped state than what is actually observed. However, this calculation rests 
on the assumption that the interaction $V$ does not depend on $x$, whereas in 
reality (if, for example, AF spin fluctuations are considered) $V$ presumably has to be reduced in parallel to \nn. We have indicated this by the thin-broken line in 
Fig.\ref{fig:pd}, which shows $T_c$ assuming $V$$\propto$$(n/n_{\rm SC})^2$, as one would expect for an interaction mediated by spin fluctuations.  

We also want to mention that $\tilde T$ is surprisingly large even for $x$=0.05, where it is about half the value of $T^{SC}_{c}$. Detecting $\tilde T$ 
may not be so easy, however. The excitation spectrum does not change appreciably upon crossing $\tilde T$ - the FS along the BZ, already established, remains 
unaffected, whereas along the anti-nodal direction the proximity-induced gap in all likelihood suffices to prevent the creation of a FS, which would 
otherwise occur. At best, one can expect a small increase in spectral weight at $E_{\rm F}$ as this temperature barrier is crossed. The temperature regime 
between $\tilde T$ and $T_c$ (SC/F) can with justification be called a ``fluctuating'' regime. This is terminology borrowed from recent investigations into the 
Nernst effect in LSCO and YBCO \cite{Wang_2, Rullier_1}, leading to the establishment of an additional temperature $T^\nu$, below which vortex-like excitations 
can be observed, and thus local (preformed) superconductivity may exist. Although $T^\nu$ has the same doping dependence as $\tilde T$ (as is found here), 
it reaches more than 
100K in optimally doped LSCO, in apparent disagreement with the phase diagram above. Nevertheless, the question of an exact definition of $T^\nu$ is open as is a 
full investigation of the effects of magnetic fields on mixed AF/SC regimes still lacking. The phase diagram of Fig.\ref{fig:pd} is also in remarkable agreement with 
one proposed recently for La-doped Bi-2212\cite{Oh_1}. After a careful analysis of resistance data, there the authors demonstrated that in the regime 
immediately above $T_c$, 
the resistance can be described via a simple universal two-component scaling function, one component with an insulator-like weak temperature dependence 
and comparably large resistance (i.e., the hole-poor phase), and a second metallic one with small resistance and related (by the authors) to SC fluctuations. 
This is precisely the scenario described here. Resistor network calculations similar to those sometimes used in manganites may help to confirm these 
results for the model considered here. NMR investigations for underdoped LSC0\cite{Singer_1} paint a very similar picture by demonstrating that the hole density $x_{\bf i}$ 
varies on a nm length scale. The reported standard deviations in the local charge density, $\Delta$$x_{\bf i}$, are comparable with the ones found 
here if both the temperature and the assumptions of model are taken into account. Finally, we also want to mention 
here that the picture emerging for the SC part of the underdoped phase is closely related to the phase-separation scenario proposed by Emery and Kivelson more than 
a decade ago\cite{Emery_1}. Furthermore, the existence of more than one characteristic temperature above $T_c$ in the underdoped regime was 
recently discussed by Lee {\it et al}.\cite{Lee_1}.      
  
Naturally, $T^\star_{\rm SC}$$\equiv$$T_c$ in the overdoped limit, since it is regarded as a homogeneous system, governed by conventional BCS physics.  
Interestingly, the fact that $p_{\rm c}$ is larger than 50$\%$ immediately suggests the opening of a ``glassy'' phase without LRO, 
which will occur once AF and SC volume fractions are approximately balanced. 
If quantum fluctuations were involved, they would presumably tend to increase the percolation threshold, i.e., stabilize the phase without LRO. 
Yet, this should only be a minor effect and leave the generic phase diagram unchanged. Nevertheless, due to the discussed proximity effects, an SC/AF coexisting 
phase may appear as well and is actually present and indicated in Fig.\ref{fig:pd}. 

If one {\it assumes} that doping in cuprates in general is facilitated by introducing hole-rich clusters with (in a first approximation) $x_{\rm c}$, 
than $x_{\rm AF}$ will be proportional to $x_{\rm c}$. In fact, the fairly quick disappearance 
of the AF phase is a logical consequence, for it is caused by the comparatively large size of the metallic/SC clusters, and which invalidates the naive 
expectation of an AF percolation threshold $x_{\rm AF}$$\sim$30$\%$-50$\%$ \cite{comment_stripe}. 
The key issue here is that it is the {\it partial delocalization of holes} that is 
responsible for the much quicker than anticipated loss of antiferromagnetism.  
For example, for the realistic value $x_{\rm c}$$\sim$0.15 (rather than $x_{\rm c}$ used here) and assuming that antiferromagnetism vanishes according 
to the rules of classical percolation, $x_{\rm AF}$=0.06, a value close to the experimentally realized one. In parallel, the onset of SC order is predicted 
for $x_{\rm SC}$=0.09, with a glassy regime in between.     
Nevertheless, these considerations only hold for a truly 2D system; if, for example, a bilayer material such as Bi-2212 is considered, the 
transition into the SC state will take place at smaller doping values since the percolation threshold is reduced in 2D when compared to 
layered or even truly 3D systems, assuming here that the value of $x_{\rm c}$ is not affected by the lattice geometry. This consideration also applies 
to the AF phase, and thus bi- and tri-layered materials should almost always prefer a SC/AF coexisting regime rather than the spin-glass phase 
encountered in single-layer materials.
        
Finally, we wish to comment on another subtle issue. The reader may have noticed that our implementation of the doping process is not entirely consistent. 
If doping is really a random process, and there are no indications otherwise, the charge-depleted volume fraction will not increase linearily with doping. 
Rather, clusters will sooner or later start to overlap, and thus the increase in metallic volume will be slower than linear. Consequently, the density 
of holes in those areas will increase somewhat from the original level, here assumed as the one of optimal doping, because the holes are squeezed into 
a relatively smaller volume. Therfore, on should assume that initially the hole-density of metallic neighborhoods is {\it smaller} than $x_{\rm c}$. 
If, for example, originally the charge density of metallic clusters were $n_{\rm SC}$$\sim$0.90, then this would in parallel 
shift the breakdown of antiferromagnetism to $x_{\rm AF}$=0.04 rather then the value of 0.06 introduced above \cite{comment_onset}. 
$x_{\rm c}$ would still be the doping level upon where 
the system becomes homogeneous, which has to happen eventually, and the relation between $x_{\rm AF}$ and $x_{\rm c}$ could be figured out by geometric considerations, 
for example by covering the lattice with randomly placed circles of a certain radius, and determining when the underlying lattice is fully covered\cite{comment_frac}. 
The decrease of density in hole-rich cluster is then reflected in the observed small downward-shift of the chemical potential {\it within the MH gap}, 
which was remarked upon in Section III and observed experimentally\cite{Shen_1}. 
\section{Conclusions}
Motivated by a series of recent PE experiments we have explored the physics of underdoped cuprates under the assumption that they represent 
a mixture of the undoped parent insulator and a SC/metallic phase with a hole-density similar to the one of optimal doping. 
As emphasized by us previously\cite{Mayr_1}, this scenario is consistent with most of the PE data, specifically the two-branched signals that are observed 
throughout the underdoped regime, as well as the development of the nodal FS arc with doping. Disorder as it manifests itself in spatially varying chemical 
potentials seems to be the driving force in LSCO, but inhomogeneity is not necessarily tied to disorder alone. Here, we point out the importance of 
proximity effects for such exotic electronic states, and how they can explain even subtle details of experimental results, such as the development of the 
low-energy excitation gap in the anti-nodal direction and the behavior of the spectral weight at $E_{\rm F}$. Proximity effects are always present in 
inhomogeneous electronic systems, although they may be hidden if both competing orders are of sufficiently equal strength. We have also extended our 
previous calculations to finite temperatures and predict a gradual increase of the FS arc for temperatures above $T_c$ and that the intensity along 
the BZ diagonal should remain constant, both in principle verifiable directly by ARPES measurements. 

In the underdoped phase, doping is tantamount to inserting clusters of hole density $x$$\leq$$x_{\rm c}$ 
into the insulating host. The accompanying phase transitions are percolative in nature, irrespective of whether they are caused 
by a change in temperature or doping. Pseudogaps appear as hallmarks of remnant antiferromagnetism and mostly arise because of the vast difference in 
energy between SC and AF excitations. Transport properties aside, the superconductor-metal transition for $x$$<$$x_{\rm c}$ is a smooth one, quite 
unlike what is observed for overdoped materials, where it is marked by the sudden 
appearance of Bogoliubov quasiparticles in both experiment and theory. Above $T_c$, the underdoped phase possesses isolated SC clusters, 
which may survive even into the AF phase and up to surprisingly large temperatures. This gives rise to an additional characteristic temperature 
$\tilde T$, as independently suggested by Nernst effect and resistance measurements. In addition, the variation in local hole density is approximately as 
large as the one recently reported in LSCO.  
 
Overall, it was shown that those fairly simple considerations can explain a substantial part of the high-$T_c$ phenomenology, including the phase diagram. 
Moreover, no further assumptions of competing states, changes in parameters, curious non-Fermi-liquid metallic states, etc. have to be made to 
rationalize an enormous variety of experimental data. From this point of view, the most important task in cuprate theory is to decipher the effects 
of chemical doping, particularly with respect to possible spatial chemical potential fluctuations. 

We also want to stress here that similar ideas have been widely discussed in manganites \cite{Dagotto_1}, and recent ARPES data on the doped bilayer 
material La$_{2-2x}$Sr$_{1+x}$Mn$_2$O$_7$ are remarkably similar to those of underdoped cuprates \cite{Mannella_1}.  
Most importantly, a two-peak structure in spectral intensity was found along the BZ diagonal (though not along the anti-nodal direction) for $x$=0.4. 
Since we have shown that such features are inevitable consequences of mixed-phase states, 
it seems natural to interpret those data along the same lines, the only difference being the competing phases in question. 
In manganites, one of the phases is the ferromagnetic metal, whereas the competing insulating one is either a CO one and/or an AF insulator. 
Nevertheless, given what we know about the cuprates, this explanation seems more reasonable 
and natural than the one suggested by the authors of Ref.\onlinecite{Mannella_1}, which relied on strong electron-phonon coupling, 
especially since the evidence for a phase mixture is stronger in manganites than in cuprates to begin with \cite{Dagotto_1}. 
The underlying theme of emerging complexity in both cuprates and manganites was recently discussed in Ref.\onlinecite{Dagotto_2}.
  
In this context, it would be particularly interesting to obtain ARPES results for slightly doped La$_{1-x}$Sr$_x$MnO$_3$, since this material is most closely 
related to LSCO and has a comparatively simple phase diagram, with only the half-filled Mott insulator and the ferromagnetic metallic phase present for 
doping levels smaller than $x$=0.5. If the mixed-state picture applies there as well, the resulting PE spectra will look just 
like the ones in LSCO, including a PG as postulated earlier\cite{Moreo_1}. In particular, based on the proximity effects emphasized in this work, 
one would expect an excitation gap in the anti-nodal direction even 
though the ferromagnetic metallic phase is nominally ungapped, unlike the SC one. From what we know about spectral functions in electronically inhomogeneous 
systems, such experimental results, if they became available, should be regarded as the best indicator of phase separation in slightly doped manganites, 
ruling out the existence of a canted phase.

\begin{acknowledgments}
Discussions with Elbio Dagotto are gratefully acknowledged. This work was supported by NSF grant DMR-0443144.
\end{acknowledgments}
\bibliographystyle{unsrt}

\begin{thebibliography}{20}
\bibitem{Comment_1} The smaller $J$, the lesser is the energy penalty for broken AF bonds; the charge carriers spread out further, and consequently the hole 
density (per site) is reduced. Therefore, the charge density at charge-depleted sites is larger for smaller values of $J$.
\bibitem{comment_stripe} There is another potentially interesting point to consider: we have assumed rotational invariance of the metallic clusters. Nevertheless, due to the 
tendency to stripe-formation, those cluster may prefer to be elongated in one direction, which may change $p_{\rm c}$ to some degree.
\bibitem{comment_onset} And, vice versa, this (together with proximity effects) also would explain why the onset of superconductivity is observed for values of 
$x$ smaller than $x_{\rm SC}$=0.09.
\bibitem{comment_frac} Preliminary calculations show that if a single circle covers 10$\%$ of the underlying lattice, it takes on average 30 such circles 
(instead of $\sim$10) to cover 95$\%$ in case their positions are uncorrelated. The coresponding number for circles with twice the area is 15. Note, however, that 
the approximation of independent circles is not fully accurate since charged impurities will try to avoid each other to some degree.
\end{thebibliography}

\end{document}